\documentclass[aps,pra,twocolumn,graphics,epsfig,floats]{revtex4}
\usepackage{graphicx}

\newcommand{\be}{\begin{eqnarray}}
\newcommand{\ee}{\end{eqnarray}}

\begin{document}

\title{Could Energy Decoherence due to Quantum Gravity be observed?}

\author{Christoph Simon$^{1}$ and Dieter Jaksch$^{2}$}
\address {$^1$ Laboratoire de Spectrom\'{e}trie Physique, CNRS et Universit\'{e} Joseph Fourier Grenoble-I, F-38402 St. Martin
d'H\`{e}res, France\\  $^2$ Clarendon Laboratory, University of
Oxford, Parks Road, Oxford OX1 3PU, United Kingdom}

\date{\today}

\begin{abstract}
It has recently been proposed that quantum gravity might lead to
the decoherence of superpositions in energy, corresponding to a
discretization of time at the Planck scale. At first sight the
proposal seems amenable to experimental verification with methods
from quantum optics and atomic physics. However, we argue that the
predicted decoherence is unobservable in such experiments if it
acts globally on the whole experimental setup. This is related to
the unobservability of the global phase in interference. We also
show how local energy decoherence, which acts separately on system
and phase reference, could be detected with remarkable sensitivity
and over a wide range of length scales by long-distance Ramsey
interferometry with metastable atomic states. The sensitivity of
the experiments can be further enhanced using multi-atom
entanglement.
\end{abstract}

\maketitle

\section{Introduction} \label{intro}

The unification of quantum physics and gravitation is one of the
big open questions in physics. A large amount of theoretical work
is devoted to it, following various approaches, including string
theory, loop quantum gravity and others. Experimental guidance
would be extremely valuable. Some predictions of string theory,
such as supersymmetry, should be testable in future high-energy
particle accelerators. Certain candidate theories predict
deviations from the usual dispersion relations for very energetic
particles. Astrophysical observations can give constraints on such
predictions \cite{amelinocamelia}.

Another class of theoretical predictions from quantum gravity
concerns the decoherence of quantum superpositions
\cite{ellis,penrose}. In particular it has recently been predicted
based on discrete quantum gravity that there should be decoherence
in the energy basis \cite{pullin}. Such energy decoherence can be
understood intuitively as arising from a discretization of time at
a very small scale \cite{milburn}. Even though there is no
universal agreement that quantum gravity implies energy
decoherence, it seems to us that the proposal deserves serious
attention, not only because it is related to a very basic concept
(the discretization of time), but also because at first sight it
seems that it might lend itself to experimental verification with
the methods of quantum optics and atomic physics, i.e. essentially
tabletop experiments \cite{pullin}.

Ref. \cite{pullin} predicts that the evolution equation for the
density matrix $\rho$ of a quantum system with Hamiltonian $H$
should be given by
\begin{equation}
\frac{d\rho}{dt}=\frac{i}{\hbar}[H,\rho]- \frac{\sigma}{\hbar^2}
[H,[H,\rho]], \label{mastereq}
\end{equation}
where $\sigma$, which is essentially the discretization timescale,
may be as small as the Planck time $t_P$, i.e. of order
$10^{-43}$s. This time evolution leads to a decay of off-diagonal
terms in the energy basis $|E\rangle \langle E+\Delta E|$ with a
rate of order $\sigma (\Delta E/\hbar)^2$.

In this paper we address the question whether this type of
decoherence could be observed in practice. We believe that the
only presently conceivable type of experiment that has the
potential to yield non-trivial bounds on the parameter $\sigma$ of
Eq. (1) is to prepare a superposition of two states with
substantially different energies and observe its decoherence,
trying to separate environmental and potentially present quantum
gravitational effects. The quadratic dependence on $\Delta E$ of
the decay rate for such superpositions is a decisive advantage
compared to effects for wavepackets with a smooth distribution in
energy, which would only become significant for energies close to
the Planck energy, cf. Ref. \cite{milburn}.

The most promising approach to achieve such superpositions is the
use of long-lived metastable atomic states that are separated from
the ground state by optical transitions, such as the $^3P_2$ and
$^3P_0$ states in Strontium \cite{katori}. A superposition
$|g\rangle+|e\rangle$ between the ground state and such an excited
state has a $\Delta E$ of order 1 eV. Note that this is about
seven orders of magnitude larger than, for example, the
cavity-induced energy splitting discussed in \cite{pullin} would
be in a realistic experiment \cite{kimble}. Moreover, because the
spontaneous lifetime of the excited state can be of the order of
hundreds of seconds \cite{katori}, an experiment with a single
atom could in principle be sensitive to a decoherence rate at the
level of $10^{-3}/$s. This can again be compared for illustration
to the example discussed in Ref. \cite{pullin}, where typical
cavity lifetimes are of order microseconds \cite{kimble}. Such
sensitivity to decoherence rates in the mHz regime would allow one
to detect $\sigma$ at the level of $10^{-33}$s, corresponding to
an energy scale of $10^{18}$ eV, far beyond the energies of
current or projected particle accelerators, which are at the level
of $10^{12}$ to $10^{13}$ eV. This remarkable sensitivity could be
further improved by several orders of magnitude using multi-atom
entangled states of the GHZ type, $|e\rangle^{\otimes
N}+|g\rangle^{\otimes N}$. We will discuss the possibility of
using multi-atom entanglement in more detail below.

However, we first have to point out a fundamental problem
concerning the observability of {\it global} energy decoherence in
the quantum optical domain. Coherence in energy (and therefore
also decoherence) is in practice observed interferometrically by
studying the {\it phase} of the quantum system under consideration
{\it relative} to a given phase reference. Only the relative phase
between system and reference is observable. At the fundamental
level the phase reference also has to be treated as a quantum
system. Energy decoherence that acts globally on system and
reference together only has an effect on the global phase of the
combined system, but does not influence the relative phase between
the two parts. It is therefore unobservable. We discuss this
rather subtle point in detail in section \ref{unobservability}
with the help of illustrative examples. This part of our work can
be seen as an extension of the arguments by Finkelstein
\cite{finkelstein} to realistic experimental situations. We
conclude this section by arguing that the need for a phase
reference, and thus the unobservability of global decoherence,
fundamentally arises from the fact that macroscopic systems have
very little coherence in energy.

In contrast to global decoherence, {\it local} energy decoherence
that acts separately on the system and the reference can be
observable. In section \ref{tests} we describe several
experimental approaches that might allow to test energy
decoherence which is local on various length scales, ranging from
micrometers to possibly millions of kilometers.

\section{Unobservability of global energy decoherence}
\label{unobservability}

\subsection{Ramsey Interference}

We will first illustrate the problem by discussing a single atom
that can be in states $|g\rangle$ or $|e\rangle$, but we will
argue below that it is much more general. It is important to
consider how a superposition of the type $|g\rangle+|e\rangle$
would in practice be created and observed, namely by applying
laser pulses that induce Rabi rotations between $|g\rangle$ and
$|e\rangle$. The experimental technique is known as Ramsey
interferometry. The first pulse creates the superposition
$|g\rangle+|e\rangle$ from the initial state $|g\rangle$, followed
by a variable delay, during which the two terms in the
superposition acquire a relative phase, such that the state
evolves to $|g\rangle+e^{i\phi}|e\rangle$, and during which
decoherence can act. The second pulse is such that it brings the
system back to $|g\rangle$ with unit amplitude if the relative
phase $\phi$ is zero and if no decoherence has occurred. If $\phi$
is different from zero, the probability to observe the system in
$|g\rangle$ will be different from one, leading to Ramsey
interference fringes in dependence on the phase. Energy
decoherence as described by Eq. (1) will affect the visibility of
these fringes. In particular, for complete decoherence the
superposition is transformed into an equal mixture
$\frac{1}{2}(|g\rangle\langle g|+|e\rangle \langle e|)$, in which
all phase information is lost and which is unchanged by the second
laser pulse. Note that at the end the system is always detected in
either $|e\rangle$ or $|g\rangle$, i.e. the detection is performed
in the basis spanned by the energy eigenstates, not in a basis of
superposition states. This detection is typically performed by
detecting fluorescence (under laser excitation) from a third state
that is accessible only from one of the two states.

Ramsey interference occurs because there are two different
histories that can lead to the same final state. Suppose that at
the end the atom is detected in $|e\rangle$. Then it can have
absorbed a photon from the first pulse and acquired a phase $\phi$
during the intermediate waiting time, or it can have absorbed a
photon from the second pulse and thus not acquired a phase. The
amplitudes for the two histories have to be added. The two
histories are distinguished by the energy of the atom during the
waiting time. It therefore seems at first sight that the
experiment should be sensitive to decoherence in the energy basis.

Conventionally in the description of Ramsey interference the laser
light is treated as a classical system, which is usually an
excellent approximation. However, for our present purpose it is
essential to take the quantum character of the light into
consideration. The Hamiltonian for the interaction between the
atom and the light is
\begin{equation}
H=g(a |e\rangle \langle g|+a^{\dagger} |g\rangle \langle e|).
\end{equation}
This is in the rotating wave approximation, which is extremely
well justified in the relevant regime where the pulse durations
are much longer than an optical period. This Hamiltonian describes
the exchange of excitations between the light and the atom. The
destruction of a photon is accompanied by the creation of an
atomic excitation and vice versa. The total energy of the combined
system light plus atom is not changed by the action of this
interaction Hamiltonian.

The Ramsey interference can be analyzed independently in subspaces
of fixed total energy, that is for initial states $|N\rangle
|g\rangle$, where $N$ is the total number of photons. Let us first
consider an idealized situation where the same laser pulse is made
to interact twice with the atom. The first interaction creates a
superposition of the two terms $|N\rangle |g\rangle$ and
$|N-1\rangle |e\rangle$. Both terms can give rise to the states
$|N\rangle |g\rangle$ and $|N-1\rangle |e\rangle$ after the second
interaction. The relative phase that the two terms acquire between
the interactions therefore determines the observable interference
effects (the probabilities for the atom to be in $|g\rangle$ or
$|e\rangle$). Obviously the two terms have the same total energy.
The observable effects are therefore completely independent of the
presence of global energy decoherence, that is of energy
decoherence that acts on the light and the atom as on a single
system.

It is always possible to analyze the interference in this way,
even if the light is not initially prepared in a Fock state
$|N\rangle$. In particular, this is also true for the case where
the state of the light is a macroscopic coherent state, such that
it remains essentially unchanged by the exchange of photons with
the atom, and always factorized from the atomic state. Our
statement here is not that a true superposition state of the atom
can never be created, which is true for an initial Fock state of
the light field. The essential point is that in any case only the
interference within each pair of histories with the same total
energy matters for the observable effects.

The same conclusion can be reached in a more realistic situation,
where the light is split into two pulses, that interact
sequentially with the same atom. Again the analysis can be
performed in subspaces of fixed total energy. The initial $N$
photon state is split coherently into two parts, corresponding to
the first and the second pulse. This creates a state of the form
$\sum_n c_n |N-n\rangle_1 |n\rangle_2 |g\rangle$. The first pulse
interacts with the atom, such that every component $|N-n\rangle_1
|g\rangle$ is transformed into a superposition $a_n |N-n\rangle_1
|g\rangle + b_n |N-n-1\rangle_1 |e\rangle$, where the interaction
is adjusted such that the coefficients $a_n$ and $b_n$ are equal
to $1/\sqrt{2}$ with good approximation. After this interaction
the first pulse never comes back. Therefore it does not contribute
to the interference. The global state after the interaction
between the first pulse and the atom can be rewritten as
\begin{equation} \sum_n |N-n\rangle_1 \left( c_n a_n |n\rangle_2
|g\rangle + c_{n-1} b_{n-1} |n-1\rangle_2 |e\rangle \right).
\end{equation}
Tracing over the first pulse, one sees that the relevant
interference is between pairs of terms of the form $|n\rangle_2
|g\rangle$ and $|n-1\rangle_2 |e\rangle$, that is between pairs of
states with the same energy. Both of these terms can lead to final
states $|n\rangle_2 |g\rangle$ and $|n-1\rangle_2 |e\rangle$ after
the interaction between the second pulse and the atom. Therefore
the phase between them determines the final probability for the
atom to be detected in $|g\rangle$ or $|e\rangle$. Again the
relevant interference is between states with the same total
energy, and the final probability for the atom to be in
$|g\rangle$ or $|e\rangle$ is completely independent of energy
decoherence that acts on the second pulse and the atom as on a
single system.

This unobservability of global energy decoherence is not specific
to Ramsey interference. We argue that it is universal, at least
for the domain of quantum optics. The basic reason is that in
every experiment conceivable to us in the quantum optical regime
the final detection is performed in the energy basis.
Superpositions in energy, such as the fixed phase relationship
between $|g\rangle$ and $|e\rangle$, are detected with the help of
a phase reference. In Ramsey interference the superposition is
created by the first laser pulse, while the second laser pulse
serves as the phase reference. What matters for the experimental
results is the relative phase between system and the reference.
The global phase is not observable. However, it is only this
global phase that is affected by global energy decoherence.

\subsection{Michelson Interference}

To further emphasize and clarify this important point, consider a
very simple example of such an interference experiment, a
Michelson interferometer for light. We denote the two input modes
of the interferometer by $a$ and $b$, and the modes traveling
towards the mirrors in the two arms by $c$ and $d$. We start with
a coherent state $|\alpha\rangle$ in mode $a$ and the vacuum in
mode $b$, i.e. an initial state
\begin{equation}
|\alpha\rangle_a|0\rangle_b=e^{-|\alpha|^2/2}\sum
\limits_{n=0}^{\infty}
\frac{\alpha^n}{\sqrt{n!}}|n\rangle_a|0\rangle_b=e^{-|\alpha|^2/2}e^{\alpha
a^{\dagger}}|0\rangle, \end{equation} where $|0\rangle$ is the
vacuum of all modes. The modes $c$ and $d$ are related to the
inputs by $a =(c+d)/\sqrt{2}$ and $b=(c-d)/\sqrt{2}$, such that
\begin{equation} |\alpha\rangle_a
|0\rangle_b=|\frac{\alpha}{\sqrt{2}}\rangle_c
|\frac{\alpha}{\sqrt{2}}\rangle_d. \end{equation} The incoming
light is split equally between the two arms. The modes $c$ and $d$
are transformed into modes $\tilde{c}$ and $\tilde{d}$
respectively, which travel from the mirrors back to the
beamsplitter. For a perfectly balanced interferometer the output
modes $\tilde{a}$ and $\tilde{b}$ are related to $\tilde{c}$ and
$\tilde{d}$ in the same way as $a$ and $b$ are to $c$ and $d$. In
particular one therefore has $a
=(c+d)/\sqrt{2}=(\tilde{c}+\tilde{d})/\sqrt{2}=\tilde{a}$.

The coherent states propagating in the two arms are coherent
superpositions of states of different photon numbers and thus
different energies. It might therefore seem that global energy
decoherence should have an effect on the interference, such that
in the presence of decoherence some photons would end up in mode
$\tilde{b}$ in the final state. However, the effect of complete
global energy decoherence is to transform the state
$|\alpha\rangle_a|0\rangle_b$ into a Poissonian mixture of Fock
states, \begin{equation} \rho=\sum \limits_{n=0}^{\infty}
e^{-|\alpha|^2}\frac{|\alpha|^{2n}}{n!}|n\rangle_a \langle
n||0\rangle_b \langle 0|. \end{equation} This input state is
exactly reproduced in the output modes, because the interference
happens independently for every total photon number, according to
\begin{equation} (a^{\dagger})^n |0\rangle=
(\frac{c^{\dagger}+d^{\dagger}}{\sqrt{2}})^n
|0\rangle=(\frac{\tilde{c}^{\dagger}+\tilde{d}^{\dagger}}{\sqrt{2}})^n
|0\rangle=(\tilde{a}^{\dagger})^n |0\rangle. \end{equation} The
coherence (or its absence) between different total photon numbers
is irrelevant for the Michelson interference. This is in full
analogy to Ramsey interference, which, as we have seen above, can
also be analyzed separately for every total photon number.
Equivalently, the global phase of the initial state
$|\alpha\rangle$ is irrelevant, only the relative phase between
the states in the two arms is important. This relative phase
remains unaffected by the global decoherence. This can also be
seen by noting that the decohered state can be written as
\begin{equation}
\rho=\frac{1}{2\pi}\int \limits_0^{2 \pi} d\phi |\alpha
e^{i\phi}\rangle_a \langle \alpha e^{i\phi}||0\rangle_b \langle
0|,
\end{equation}
which in terms of the modes $c$ and $d$ is
\begin{equation} \frac{1}{2\pi}\int \limits_0^{2 \pi} d\phi
|\frac{\alpha}{\sqrt{2}} e^{i\phi}\rangle_c \langle
\frac{\alpha}{\sqrt{2}} e^{i\phi}| |\frac{\alpha}{\sqrt{2}}
e^{i\phi}\rangle_d \langle \frac{\alpha}{\sqrt{2}} e^{i\phi}|.
\label{after}
\end{equation}
This shows that the relative phase between the two modes is
unaffected, even though the reduced density matrix of each
individual mode ($c$ and $d$) after global energy decoherence is
given by \begin{equation} \frac{1}{2\pi} \int \limits_0^{2 \pi}
d\phi |\frac{\alpha}{\sqrt{2}} e^{i\phi}\rangle \langle
\frac{\alpha}{\sqrt{2}} e^{i\phi}|=\sum \limits_{n=0}^{\infty}
e^{-|\alpha|^2/2}\frac{|\alpha|^{2n}}{2^n n!}|n\rangle \langle n|,
\end{equation} which is a Poissonian mixture of Fock states without any phase
relation.

\subsection{Time Domain Experiments}

We claim that these simple examples are in fact generic. In
particular, the arguments apply to Ramsey type experiments that
use GHZ states of the form $|g\rangle^{\otimes
N}+|e\rangle^{\otimes N}$ instead of single atoms. They also apply
to experiments that would aim to demonstrate coherence and
decoherence in the energy basis via time measurements. For
example, one could argue that the shortness in time of a light
pulse from a mode-locked laser demonstrates coherence in energy
and that decoherence should lead to an observable broadening of
the pulse in time. However, what is really measured in practice is
the relative time between the pulse and a reference pulse. In the
simplest case, the original pulse is split into two parts, which
are recombined in a non-linear medium, where two coincident
photons (one from each pulse) can combine to give a single photon
of higher energy, which is detected. All such experiments are
sensitive only to the {\it relative} time between system and
reference, in analogy to the relative phase for the above
experiments, and thus only allow inference about the coherence in
the energy difference between system and reference, but not about
coherence in the total energy. Again, the relevant interference
occurs between states of fixed total energy, which is shared in
different ways between system and reference. This point was made
previously in a slightly different language, but also in the time
domain, by Finkelstein \cite{finkelstein}. The present point of
view is also a good way of understanding the results of the
gedanken experiments discussed very recently by Pearle
\cite{pearle} in the context of energy-driven collapse models.

\subsection{Discussion}

Global energy coherence (and thus decoherence) is unobservable in
the above experiments because the coherence is observed with
respect to a phase reference. As long as this phase reference has
to be treated as part of the quantum system that is subject to
energy decoherence, the decoherence is unobservable. If on the
other hand the decoherence acts separately on the system and the
phase reference, it can have observable effects. We will pursue
this possibility in the following section. Another theoretical
possibility allowing observation of the decoherence would be for
the phase reference to remain unaffected by the decoherence, which
would only act on the system. This would correspond to the case of
a ``completely classical'' phase reference, which could only exist
if there were somewhere a border between the quantum and the
classical world.

But what are the reasons underlying the need for a phase
reference? In a certain sense this requirement is of a practical,
not fundamental, nature. One could detect global energy
decoherence without a phase reference, if there was a macroscopic
physical system which could be in long-lived states with large
$\Delta E$ that are {\it macroscopically distinct} from states
without significant energy coherence. In this case energy
decoherence would be directly observable. However, we suspect that
no such physical system can exist in practice. Macroscopic systems
are always in contact with their environment (usually even in
thermal equilibrium). As a consequence, the states in which they
are observed have very little coherence in the energy basis. In
particular, this applies to the macroscopic systems used to
indicate measurement results in typical experiments. For instance,
in the cases considered above the measurements are done by
counting photons (from fluorescence in the case of the Ramsey
experiment, from up-conversion in the case of the short pulses,
directly from the laser in the Michelson experiment). In these
setups only energetically distinct states (absence or presence of
a photon) lend themselves to amplification to the macroscopic
level.

It is worth pointing out that the absence of energy coherence with
large $\Delta E$ for individual systems would in fact be implied
by the presence of global energy decoherence. With the age of the
universe of order $10^{10}$ years, one finds that even for
$\sigma$ from Eq. (1) of order the Planck time $t_P$ all
coherences with $\Delta E$ larger than a few meV would have
decayed. Global energy decoherence acting on the whole universe
would thus have essentially reduced it to a mixture of energy
eigenstates at the present moment. Note that this would not
necessarily have any observable consequences for experiments using
phase references or other clocks, for the reasons discussed in
this section. In a universal energy eigenstate $|E_u\rangle$,
every energy state $|E\rangle$ of an isolated subsystem is
correlated with a state $|E_u-E\rangle$ for the rest of the
universe, so that no coherence can exist in the individual system.
Global energy decoherence could thus have itself destroyed the
conditions for its observation.

\section{Testing Local Energy Decoherence} \label{tests}

In the previous section we have arrived at the conclusion that the
prediction of Ref. \cite{pullin} is likely to be untestable if the
decoherence is assumed to act globally. It is therefore important
to understand whether there is a length scale on which the
decoherence might act locally. The question of the spatial
dependence of the energy decoherence was already raised by Milburn
in Ref. \cite{milburn}, based on considerations of Lorentz
invariance. We make no attempt to answer this question here.
However, we outline some experimental approaches that might allow
to test energy decoherence that is local on various length scales.

\subsection{Atom and Molecule Interferometry}

Atom and molecule interferometry are extremely sensitive to the
occurrence of energy decoherence that acts locally on short length
scales. For example, separations of up to 20 $\mu$m between the
two paths were achieved in an interferometry experiment with
Sodium atoms \cite{kokorowski}. The rest mass of a Sodium atom
corresponds to an energy of order 20 GeV. Even for $\sigma$ in Eq.
(1) of order the Planck time, this implies a decoherence rate of
order $10^8$/s, if the decoherence acts separately on each path.
For an atom velocity of 3000 m/s as in Ref. \cite{kokorowski},
this would imply that the atoms should be significantly decohered
after propagating just 30 $\mu$m. The calculated decoherence rates
would be even more dramatic in molecule interferometry experiments
such as Ref. \cite{hackermueller}, however the achieved path
separations are only of order 1 $\mu$m. Energy decoherence acting
locally below the $\mu$m length scale is thus already ruled out by
these experiments. A further improvement of energy decoherence
bounds on short length scales could be achieved by utilizing
multi- particle entanglement enhanced atom interferometers
\cite{Dorner}.

\subsection{Long-Distance Ramsey interferometry}

As discussed above, the relevant interference in a Ramsey
experiment is between states of the form $|g\rangle |n\rangle$ and
$|e\rangle|n-1\rangle$, where $|g\rangle$ and $|e\rangle$ are
states of the atom and $|n\rangle$ and $|n-1\rangle$ refer to the
second laser pulse. Decoherence between these two states will lead
to observable effects. The length scale on which one can probe
energy decoherence is thus given by the separation between the
atom and the light during the waiting period. In the simplest
case, the light will still be inside the laser cavity during this
time, and will be switched out of the cavity at the right moment.
The important distance is then that between the laser and the
atom. This distance can be made very large in principle. For
example, it is conceivable to connect the laser to the atom by an
optical fiber which, depending on the wavelength of the light,
would allow distances of several kilometers or more. The fiber has
to be interferometrically stabilized for such an experiment, but
this seems feasible through constant monitoring of a reference
beam. Alternatively, one could consider the use of large
free-space interferometers such as those planned for gravitational
wave detection \cite{ligo}, which aim for similar distances. Much
longer distances could in principle be accessible with space-based
experiments. LISA \cite{lisa} is a project for a space
interferometer for the detection of gravitational waves. The
interferometer is basically of the Michelson type, with the beam
splitter and the mirrors located on satellites separated by $5
\times 10^9$ m. The possible waiting time for the Ramsey
experiment is limited by the laser coherence time, which can
currently be of order 1s \cite{laser}, corresponding to a distance
of $3 \times 10^8$ m. As discussed in section \ref{intro} the use
of very long-lived metastable states in such experiments should
allow one to detect a discretization of time at the level of
$10^{-33}$ s.

\subsection{Multi-atom entanglement}

The sensitivity of Ramsey type experiments could be significantly
improved by replacing the single atom with multi-atom entangled
states of the GHZ type. There have been several proposals for the
creation of atomic GHZ states \cite{atomcats,micheli,Dorner}. The
recent scheme of Ref. \cite{micheli} allows the fast creation of
approximate GHZ type states, i.e. good approximations to the state
$|e\rangle^{\otimes N}+|g\rangle^{\otimes N}$, for large numbers
of atoms $N$. The created states are superpositions of two
components centered around very different energies. The difference
in energy between the two components is of order $N \Delta E$,
where $N$ is the number of atoms and $\Delta E$ is the difference
in energy between the states $|g\rangle$ and $|e\rangle$ of a
single atom. This implies that the gravitational decoherence rate
in such a state will be enhanced by a factor of $N^2$ compared to
a single-atom superposition.

In a generalized Ramsey type experiment one would first create the
large superposition state by letting the atoms interact for a
certain time in the presence of a laser beam in resonance with the
relevant transition \cite{micheli}, followed by a waiting period
during which the decoherence could act on the state. Then the
laser beam and the interaction would be turned on again (cf.
below), leading to a partial revival, whose amplitude would allow
one to deduce the amount of decoherence. There are several other
decoherence processes in such a scheme whose effects would have to
be distinguished experimentally from the quantum gravitational
decoherence, in particular atom losses due to spontaneous emission
and to inelastic collisions. The loss of a single particle
destroys the GHZ type superposition state. The presence of these
processes determines the in principle achievable sensitivity for
energy decoherence. Discrimination of the different decoherence
processes is facilitated by the fact that they scale differently
with the particle number and the volume of the system. However,
for simplicity and safety we will here assume that the
gravitational decoherence rate has to be larger than all other
decoherence rates in order for a clear experimental detection to
be possible.

We will here discuss the example of Strontium, with the atomic
ground state $^1S_0$ as $|g\rangle$ and the metastable $^3 P_0$
state as $|e\rangle$. The advantage of $^3 P_0$ compared to $^3
P_2$ is that inelastic two-body collisions should be strongly
suppressed because the inelastic loss channels studied in
\cite{greene} are absent for the singlet states. It should be
possible to create a Strontium BEC in either $^3 P_0$ or $^1S_0$
by optical cooling \cite{barrett}, and also to trap both states
simultaneously, as required for the present experiment, using far
off-resonant optical traps. The scheme for creating GHZ type
superposition states is described in detail in Ref.
\cite{micheli}. What is important for us here is the timescale on
which the superposition is created, which is of order $1/(N
\chi)$, where $N$ is the number of atoms in the BEC and
\begin{equation}\chi=\frac{2 \pi \hbar}{m V}(a_{gg}+a_{ee}-2
a_{eg}).\label{chi}\end{equation} Here $m$ is the mass of an atom,
$V$ is the volume of the BEC and the $a_{ij}$ are the elastic
scattering lengths for collisions between two atoms in
$|g\rangle$, between two atoms in $|e\rangle$, and between one
atom in $|e\rangle$ and one in $|g\rangle$ respectively. For later
convenience we define the coefficient $\kappa=2 \pi
\hbar/m(a_{gg}+a_{ee}-2 a_{eg})$, such that $\chi=\kappa/V$. After
creating the macroscopic superposition state, the laser coupling
the ground and excited states is turned off, and the parameter
$\chi$ is tuned to $\chi=0$, freezing the dynamics. This can be
achieved by changing the relative magnitudes of the scattering
lengths using e.g.~an optical Feshbach resonance \cite{Fesh}.
After a variable waiting period, the laser beam and $\chi$ can be
turned on again to induce a revival as detailed in Ref.
\cite{micheli}.

For the experiment to be feasible, the time for creating the
superposition has to be shorter than all relevant decoherence
times. Otherwise decoherence during the creation process would
prevent the superposition state from being formed. Moreover, the
decoherence rate due to quantum gravity should be comparable to
the decoherence due to particle loss. As stated above, we will
here assume that it has to be larger. The relevant losses are due
to spontaneous emission and to inelastic three-body collisions.
The loss rate from spontaneous emission is given by $N \Gamma$,
where $\Gamma$ is the spontaneous decay rate of the metastable
state. The three-body loss rate is of the form $k_3 N^3/V^2$,
where $k_3$ is the three-body loss coefficient. Finally the
quantum gravitational decoherence rate that we want to detect is
of the form $\gamma N^2$, where we have defined $\gamma=\sigma
(\Delta E/\hbar)^2$.

The requirement that the creation of the superposition has to be
faster than the gravitational decoherence gives  $\kappa/(NV) >
\gamma$, while the requirement that the gravitational decoherence
should dominate the other decoherence processes gives the
conditions $\gamma > \Gamma/N$ and  $\gamma
> k_3 N/V^2$. Combining the first and second of these three
inequalities gives $V < \kappa/\Gamma$. One should choose $V$ not
much smaller than this limit, in order to keep the three-body
losses as small as possible. The detectable level of gravitational
decoherence is then determined by the two conditions $\gamma >
\Gamma/N$ and $\gamma > k_3 \Gamma^2 N/\kappa^2$, where the first
bound varies as $1/N$ and the second one as $N$. This implies that
the smallest possible value for $\gamma$ is attained for $N$ of
order $\kappa/\sqrt{k_3 \Gamma}$. The minimum detectable $\gamma$
is then of order $\sqrt{\Gamma^3 k_3}/\kappa$.

The values of the above quantities can be estimated in the
following way. The spontaneous decay rate $\Gamma$ is of order
$10^{-3}$/s for the extremely long-lived $^3 P_0$ state
\cite{3P0}, and the energy separation $\Delta E$ is of order 1 eV.
The precise values of scattering lengths and inelastic collision
rates for Sr are unknown to the best of our knowledge. However,
based on experiments \cite{exp} and theoretical calculations
\cite{theory} for other atomic species one can obtain order of
magnitude estimates of $k_3=10^{-41}$m$^6/$s and $\kappa =
10^{-17}$m$^3$/s. For these values one finds that the optimal $N$
and $V$ are $N \approx 10^5$ and $V \approx 10^{-14}$m$^3$
respectively, which is very realistic from an experimental point
of view. The minimum detectable $\gamma$ is then approximately
$10^{-8}/$s. This corresponds to a detectable discretization
timescale $\sigma$ of order $10^{-38}$ s, five orders of magnitude
smaller than what is possible with a single atom. Let us note that
naively this timescale corresponds to an ultrahigh energy scale of
$10^{23}$ eV. These are only order of magnitude estimates, but it
is clear that the use of multi-atom entangled states promises a
dramatic improvement in sensitivity.

The multi-atom states could be integrated into a long-distance
Ramsey type experiment. The achievable distance is limited by the
decoherence rate. This leads to a trade-off between sensitivity
and accessible length scale. More quantitatively, the minimum
detectable $\gamma$ is of order $\Gamma/N$ (for the above optimal
$N$), where $\Gamma$ is the spontaneous emission rate, while the
decoherence rate is $\gamma N^2$, which is of order $\Gamma N$,
giving a length scale of $L_{max}=c/(\Gamma N)$. One therefore has
the relation $L_{max}=(c/\Gamma^2) \gamma$. Putting in the above
values, one sees that sensitivity to the minimum $\gamma$ obtained
above of order $10^{-8}/$s (and thus to $\sigma$ of order
$10^{-38}$ s) could still be achieved in an experiment spanning
thousands of kilometers.

%

\section{Conclusions}

We have argued that global energy decoherence is unobservable in
quantum optics experiments, because in practice energy coherence
and decoherence are studied by interferometry, which always relies
on the use of another system that serves as phase reference. The
fundamentally quantum mechanical character of this phase reference
is essential for our argument. The observable effects are governed
by the relative phase between these two systems, and are unchanged
by energy decoherence that acts globally on system and phase
reference together. We have suggested that the basic reason for
the need for a a phase reference (and thus for the unobservability
of global energy decoherence) is the fact that macroscopic objects
have very little energy coherence.

However, we have also shown how local energy decoherence, which
would act separately on system and phase reference, could in
principle be detected with remarkable sensitivity and over a wide
range of length scales, from micrometers to millions of
kilometers, combining methods from optics and atomic physics.
Energy decoherence acting locally below the micrometer scale is
already ruled out by atom interferometry. We hope that our present
work will provide a motivation for theoretical investigations into
the possible existence of a length scale in the predicted quantum
gravitational energy decoherence.

\begin{acknowledgements}

We thank Jorge Pullin for useful discussions. This work was
supported by the IRC network on Quantum Information Processing.

\end{acknowledgements}


\begin{references}

\bibitem{amelinocamelia} T. Jacobson, S. Liberati, and D. Mattingly, Nature {\bf 424}, 1019 (2003);
G. Amelino-Camelia, J. Ellis, N.E. Mavromatos, D.V. Nanopoulos,
and S. Sarkar, Nature {\bf 393}, 763 (1998).

\bibitem{ellis} J. Ellis, J.S. Hagelin, D.V. Nanopoulos, and M.
Srednicki, Nucl. Phys. B {\bf 241}, 381 (1984); S. Hawking, Comm.
Math. Phys. {\bf 87}, 395 (1982).

\bibitem{penrose} R. Penrose, in {\it Mathematical Physics 2000},
edited by A. Fokas {\it et al.} (Imperial College, London, 2000);
for an experimental proposal that might allow to obtain bounds on
this variety of predicted decoherence see W. Marshall, C. Simon,
R. Penrose, and D. Bouwmeester, Phys. Rev. Lett. {\bf 91}, 130401
(2003).


\bibitem{pullin} R. Gambini, R.A. Porto, and J. Pullin, Class.
Quant. Grav. {\bf 21}, L51 (2004).

\bibitem{milburn} G.J. Milburn, Phys. Rev. A {\bf 44}, 5401
(1991).

\bibitem{katori} M. Yasuda and H. Katori, Phys. Rev. Lett. {\bf
92}, 153004 (2004).

\bibitem{kimble} J. McKeever, J.R. Buck, A.D. Boozer, and H.J.
Kimble, quant-ph/0403121 (2004).

\bibitem{finkelstein} J. Finkelstein, comment on Ref. [5], Phys. Rev. A {\bf 47}, 2412
(1993); see also the reply by G.J. Milburn, Phys. Rev. A {\bf 47},
2415 (1993).

\bibitem{pearle} P. Pearle, Phys. Rev. A {\bf 69}, 042106 (2004).



\bibitem{kokorowski} D.A. Kokorowski, A.D. Cronin, T.D. Roberts,
and D.E. Pritchard, Phys. Rev. Lett. {\bf 86}, 2191 (2001).

\bibitem{hackermueller} L. Hackerm\"{u}ller, K. Hornberger, B.
Brezger, A. Zeilinger, and M. Arndt, Nature {\bf 427}, 711 (2004).



\bibitem{Dorner} U. Dorner, P. Fedichev, D. Jaksch, M.
Lewenstein, and P. Zoller, Phys. Rev. Lett. {\bf 91}, 073601
(2003).


\bibitem{laser} B.C. Young, F.C. Cruz, W.M. Itano, and J.C.
Bergquist, Phys. Rev. Lett. {\bf 82}, 3799 (1999).

\bibitem{ligo} See http://www.ligo.caltech.edu,
http://www.virgo.infn.it, and http://www.geo600.uni-hannover.de.

\bibitem{lisa} See http://lisa.jpl.nasa.gov and
http://www.srl.caltech.edu/lisa/pubs.html.

\bibitem{atomcats} J.I. Cirac, M. Lewenstein, K. Molmer, and P.
Zoller, Phys. Rev. A {\bf 57}, 1208 (1998); J. Ruostekoski, M.J.
Collett, R. Graham, and D.F. Walls, Phys. Rev. A {\bf 57}, 511
(1998); D. Gordon and C.M. Savage, Phys. Rev. A {\bf 59}, 4623
(1999); K. Molmer and A. Sorensen, Phys. Rev. Lett. {\bf 82}, 1835
(1999).

\bibitem{micheli} A. Micheli, D. Jaksch, J.I. Cirac, and P. Zoller, Phys. Rev. A {\bf 67}, 013607 (2003).


\bibitem{greene} V. Kokoouline, R. Santra, and C.H. Greene, Phys.
Rev. Lett. {\bf 90}, 253201 (2003).

\bibitem{barrett} M.D. Barrett, J.A. Sauer, and M.S. Chapman,
Phys. Rev. Lett. {\bf 87}, 010404 (2001).

\bibitem{Fesh} H. Feshbach, Ann. Phys. 5, 357 (1958); S. Inouye, M.R. Andrews, J. Stenger,
H.-J. Miesner, D.M. Stamper-Kurn, and W. Ketterle, Nature 392, 151
(1998); T. Calarco, U. Dorner, P. Julienne, C. Williams, and P.
Zoller, quant-ph/0403197.

\bibitem{3P0} See M. Takamoto and H. Katori, Phys. Rev. Lett. {\bf
91}, 223001 (2003); G. Ferrari {\it et al.}, Phys. Rev. Lett. {\bf
91}, 243002 (2003); I. Courtillot {\it et al.}, Phys. Rev. A {\bf
68}, 030501 (2003).

\bibitem{exp} E.A. Burt, R.W. Ghrist, C.J. Myatt, M.J.
Holland, E.A. Cornell, and C.E. Wieman, Phys. Rev. Lett. {\bf 79},
337 (1997); A. G\"{o}rlitz {\it et al.}, Phys. Rev. Lett. {\bf
90}, 090401 (2003).

\bibitem{theory} E. Tiesinga, S. Kotochigova, and P.S. Julienne,
Phys. Rev. A {\bf 65}, 042722 (2002); B. Bussery-Honvault, J.-M.
Launay, and R. Moszynski, Phys. Rev. A {\bf 68}, 032718 (2003).






\end{references}
\end{document}